\documentclass[twocolumn,showpacs,aps,superscriptaddress]{revtex4-1}
\include{def}
\usepackage{amsmath}
\usepackage{lineno}
\usepackage{graphicx}
\usepackage{xcolor}
\usepackage{longtable}
\usepackage{multirow}
\usepackage{makecell}

\def\be{\begin{equation}}
\def\ee{\end{equation}}

\begin{document}

\setlength\LTcapwidth{\linewidth}
\title{Geometrical scaling for light flavor hadrons}

\author{A.Lindner}
\affiliation{National Institute for Physics and Nuclear Engineering - IFIN-HH\\
Hadron Physics Department\\
Bucharest - Romania}
\affiliation{Faculty of Physics, University of Bucharest} 
\author{M.Petrovici}
\affiliation{National Institute for Physics and Nuclear Engineering - IFIN-HH\\
Hadron Physics Department\\
Bucharest - Romania}
\affiliation{Faculty of Physics, University of Bucharest}
\author{A.Pop} 
\affiliation{National Institute for Physics and Nuclear Engineering - IFIN-HH\\
Hadron Physics Department\\
Bucharest - Romania}
\date{\today}

\begin{abstract}

As it is well known by now, the pre-partonic phase in hadron collisions is successfully described by the Color Glass Condensate (CGC) approach. Previous studies, based on experimental data obtained on a wide range of energies at the Relativistic Heavy Ion Collider (RHIC) and at the Large Hadron Collider (LHC) for $\pi^+$, $K^+$ and $p$, evidenced that observables related to the dynamics of the collision, i.e. the mean transverse momentum ($\langle p_T \rangle$), the slope of the $\langle p_T \rangle$ dependence on the mass of the hadrons and the average transverse flow velocity obtained from the simultaneous fits of the $p_T$ spectra of the particles with the Boltzmann-Gibbs Blast Wave (BGBW) expression, scale rather well as a function of the square root of the ratio of the particle density over unit of rapidity to the overlapping area of the colliding nuclei ($\sqrt{(dN/dy)/S_\perp}$), the relevant scale in the gluon saturation picture. Results of a similar study extended to strange and multi-strange hadrons, for both proton-proton (\textit{pp}) and heavy-ion (\textit{A-A}) collision systems are presented in the present paper. The similarities and differences in the behaviour of strange hadrons relative to non-strange hadrons are discussed. 
\end{abstract}

\maketitle

\section{Introduction}
%The CGC theoretical approach used to describe experimental data at LHC energies, characterised by very low x values. 
The pre-partonic phase in hadron-hadron collisions at ultra-relativistic energies is successfully described by the CGC theoretical approach \cite{theonew}. 
Based on this approach and local parton-hadron duality \cite{theonew2}, predictions on the behaviour of the $\langle p_T \rangle / \sqrt{(dN/dy)/S_\perp}$ ratio as a function of collision energy and centrality were made \cite{theopredict1, theopredict2}. In the previous studies these predictions were disputed based on existing measured data. The expected behavior of $\langle p_T \rangle / \sqrt{(dN/dy)/S_\perp}$ as a function of collision energy and centrality for a wide range of energies \cite{geomscaling1} and for different collision systems at the same collision energy was systematically investigated in Ref. \cite{geomscaling2} for $\pi^+$, $K^+$ and $p$.

%%%%%%%%%%%%%%%%%%%%%%
%\section{$\sqrt{(dN/dy)/S_\perp}$ estimates}
These studies confirmed that the relevant scale in the gluon saturation picture predicted by CGC is the square root of the ratio between the particle density over unit of rapidity and the overlapping area of the colliding systems ($\sqrt{(dN/dy)/S_\perp}$). The particle densities were estimated based on available experimentally measured data \cite{geomscaling1}. For heavy-ion collisions, $S_\perp$ was estimated based on a Glauber Monte Carlo (GMC) approach by averaging the overlapping area of the nuclei over many events, as described in Ref. \cite{geomscaling1}. In the case of \textit{pp} collisions, $S_\perp$ is estimated as the overlapping area corresponding to different values of the energy density of the Yang-Mill fields, $\varepsilon = \alpha \Lambda^4_{QCD}$ ($\alpha\in$[1, 10]), based on the IP-Glasma approach \cite{Sperp_pp1, Sperp_pp2}. The $S_\perp$ values considered in the present studies correspond to $\alpha=1$ and $10$ as in Ref. \cite{geomscaling1}.

\begin{figure}[b!]
\centering
\includegraphics[width=\linewidth]{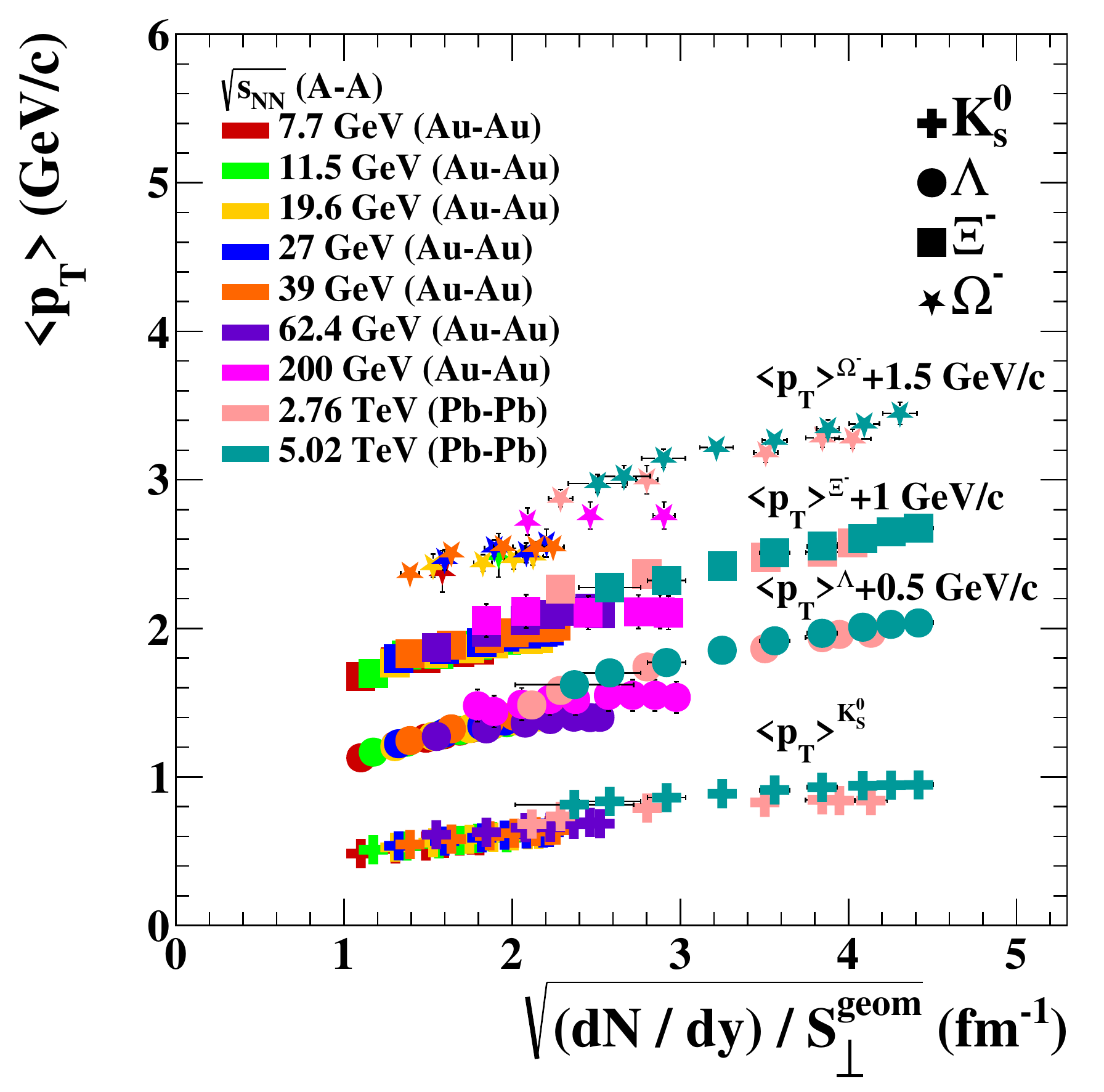}
\caption{$\langle p_T \rangle$ dependence on $\sqrt{(dN/dy)/S_\perp}$, for strange and multi-strange hadrons ($K^0_s$, $\Lambda$, $\Xi^{-}$ and $\Omega^{-}$) in heavy-ion collisions; for a better visualisation, an offset in  $\langle p_T \rangle$ was added.}
\label{fig-1}
\end{figure}

The $\langle p_T \rangle$ dependence on  $\sqrt{(dN/dy)/S_\perp}$ is presented in Fig.\ref{fig-1}, for strange and multi-strange hadrons \cite{data1}-\cite{data6}. A wide range of energies is studied, from $\sqrt{s_{NN}}$ = 7.7 GeV up to 39 GeV measured in the Beam Energy Scan (BES) program and the intermediate and highest energies measured at RHIC ( $\sqrt{s_{NN}}$ = 62.4 and 200 GeV) in \textit{Au-Au} collisions up to $\sqrt{s_{NN}}$ = 2.76 and 5.02 TeV measured at LHC for \textit{Pb-Pb} collisions. 

A very good scaling is observed from the lowest BES energy, up to the highest energy measured at RHIC, where a trend towards saturation is evidenced. With an offset between RHIC energies and the LHC energies, a very good scaling is also evidenced at the LHC energies.

\begin{figure}[t!]
\centering
\includegraphics[width=\linewidth]{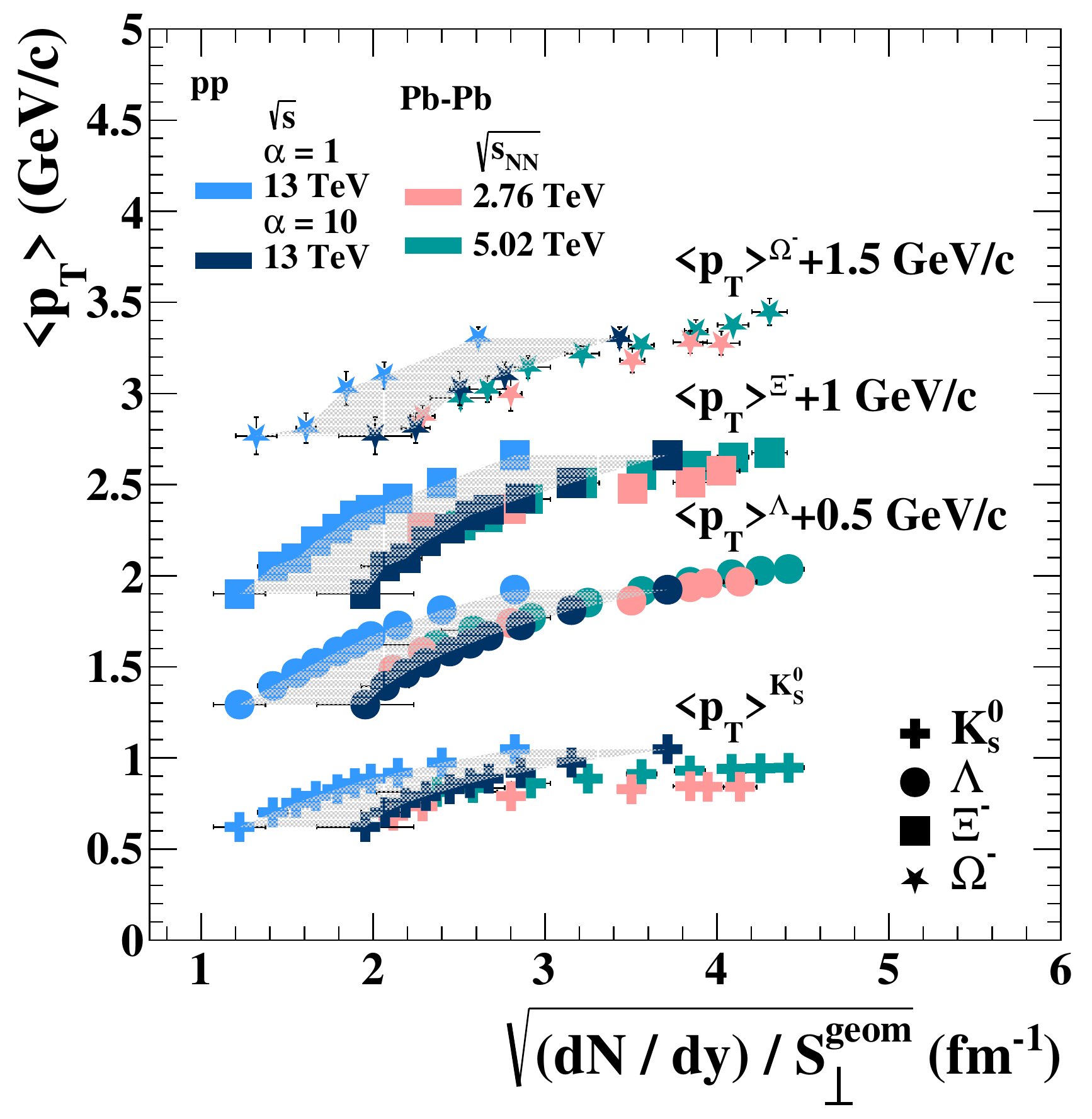}
\caption{The dependence of $\langle p_T \rangle$ on $\sqrt{(dN/dy)/S_\perp}$ for $K^0_s$, $\Lambda$, $\Xi^{-}$ and $\Omega^{-}$ in \textit{pp} and \textit{A-A} collisions at LHC energies; for a better visualisation, an offset in  $\langle p_T \rangle$ was added.}
\label{fig-2}
\end{figure}

\begin{figure}[t!]
\centering
\includegraphics[width=\linewidth]{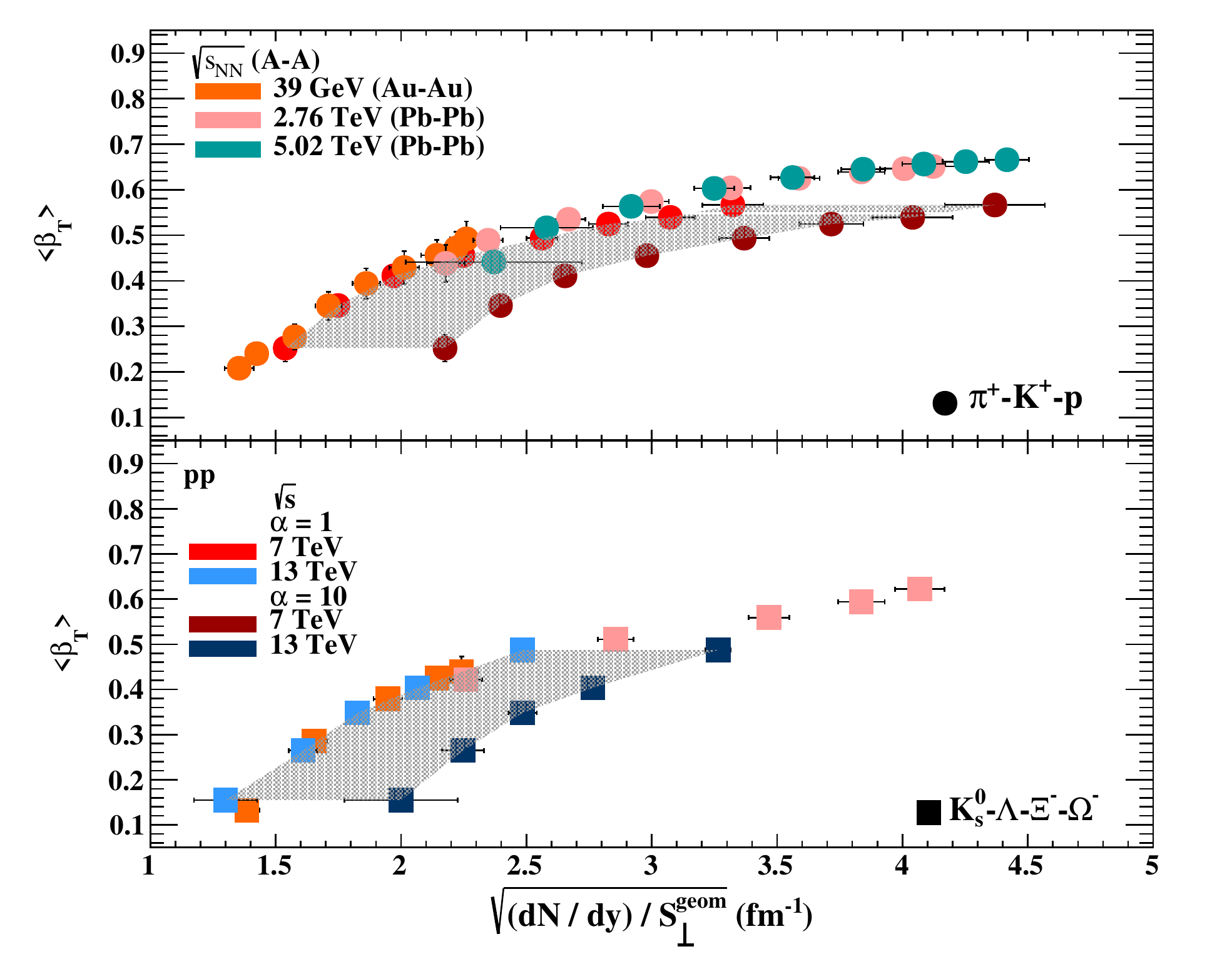}
\caption{$\sqrt{(dN/dy)/S_\perp}$ dependence of $\langle \beta_T \rangle$ obtained from the BGBW simultaneous fits of $\pi^+$, $K^+$ and $p$ (a) and  $K^0_s$, $\Lambda$, $\Xi^-$ and $\Omega^-$ (b).}
\label{fig-3}
\end{figure}

\begin{figure}[b!]
\centering
\includegraphics[width=\linewidth]{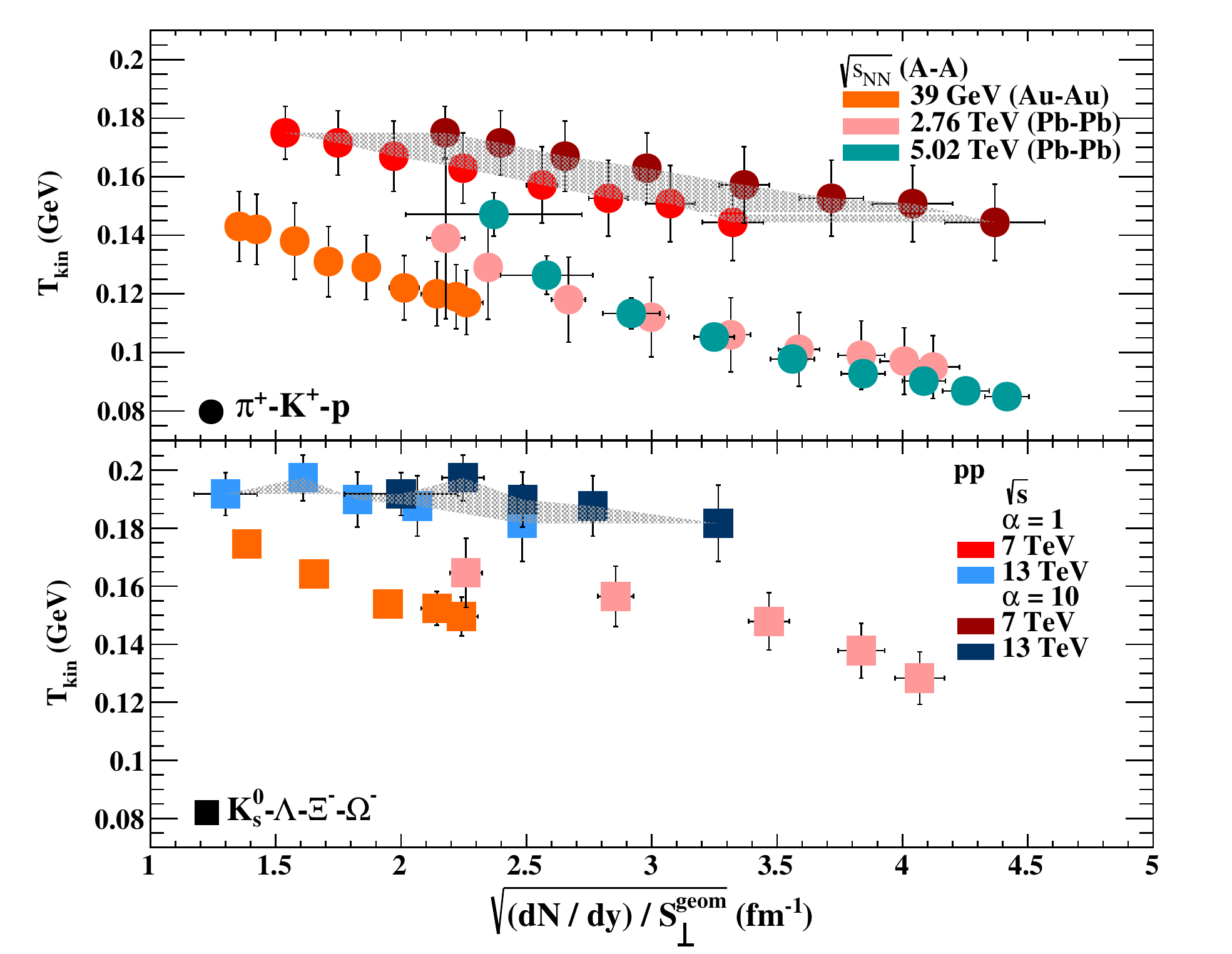}
\caption{$\sqrt{(dN/dy)/S_\perp}$ dependence of $T_{kin}$ obtained from the BGBW simultaneous fits of $\pi^+$, $K^+$and $p$ (a) and  $K^0_s$, $\Lambda$, $\Xi^-$ and $\Omega^-$ (b).}
\label{fig-4}
\end{figure}

In Fig.\ref{fig-2}, a comparison between \textit{pp} \cite{data7}-\cite{data9} and \textit{Pb-Pb} collisions at LHC energies is done in terms of the dependence of $\langle p_T \rangle$ on $\sqrt{(dN/dy)/S_\perp}$ for strange and multi-strange hadrons. A very good scaling is evidenced between \textit{Pb-Pb} and \textit{pp} collisions in the case of $\alpha=10$. The border lines correspond to the two extreme values of  $\alpha$, $1$ and $10$. Since $\alpha$ can take any value in this range, the values in-between are represented with a gray shaded area. This recipe is used for all the figures where  the results for \textit{pp} collisions are presented. A comparison between \textit{pp} and \textit{Pb-Pb} collisions in terms of the parameters obtained by a simultaneous fit of the $p_T$ spectra of different hadrons using the BGBW expression inspired from hydrodynamical models \cite{bgbw} is presented in the following. Figure \ref{fig-3} shows the dependence of the average transverse flow velocity ($\langle \beta_T \rangle$) on the geometrical variable, while the same dependence of the kinetic freeze-out temperature ($T_{kin}$) is represented in Fig.\ref{fig-4}. In Fig.\ref{fig-3}a and Fig.\ref{fig-4}a, the values of the parameters are obtained from  simultaneous fit of the $p_T$ spectra of $\pi^+$, $K^+$ and $p$ \cite{geomscaling1}, while in Fig.\ref{fig-3}b and Fig.\ref{fig-4}b, the parameters result from a simultaneous fit in the case of $K^0_s$, $\Lambda$, $\Xi^{-}$ and $\Omega^{-}$.  In Fig.\ref{fig-3}, a very good agreement is obtained between \textit{A-A} and \textit{pp} collisions (for $\alpha$=1). Higher values of $\langle \beta_T \rangle$ are evidenced in the case of non-strange hadrons. The kinetic freeze-out temperatures,  presented in Fig. \ref{fig-4}, decrease towards high particle multiplicities and centralities, showing systematically larger values for strange and multi-strange hadrons than in the case of $\pi^+$, $K^+$ and $p$ and systematically larger values for \textit{pp} at the same $\sqrt{(dN/dy)/S_\perp}$ .  

\begin{figure}[h!]
\centering
\includegraphics[width=\linewidth]{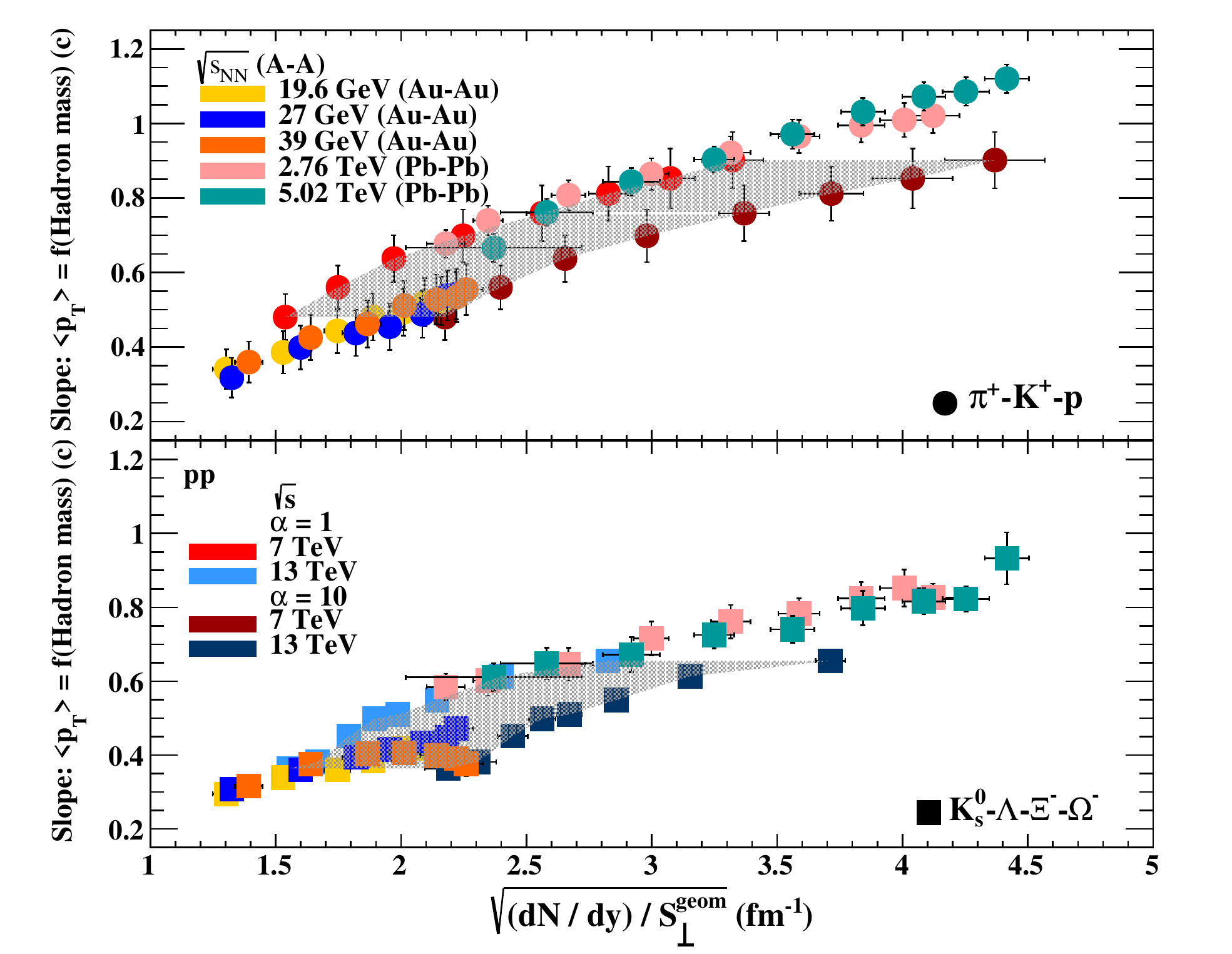}
\caption{$\sqrt{(dN/dy)/S_\perp}$ dependence of the slope obtained from the linear fit of the $\langle p_T \rangle$-hadron mass dependence for $\pi^+$,$K^+$and $p$(a) and $K^0_s$,$\Lambda$,$\Xi^-$and $\Omega^-$(b).}
\label{fig-5}
\end{figure}

\begin{figure}[b]
\centering
\includegraphics[width=\linewidth]{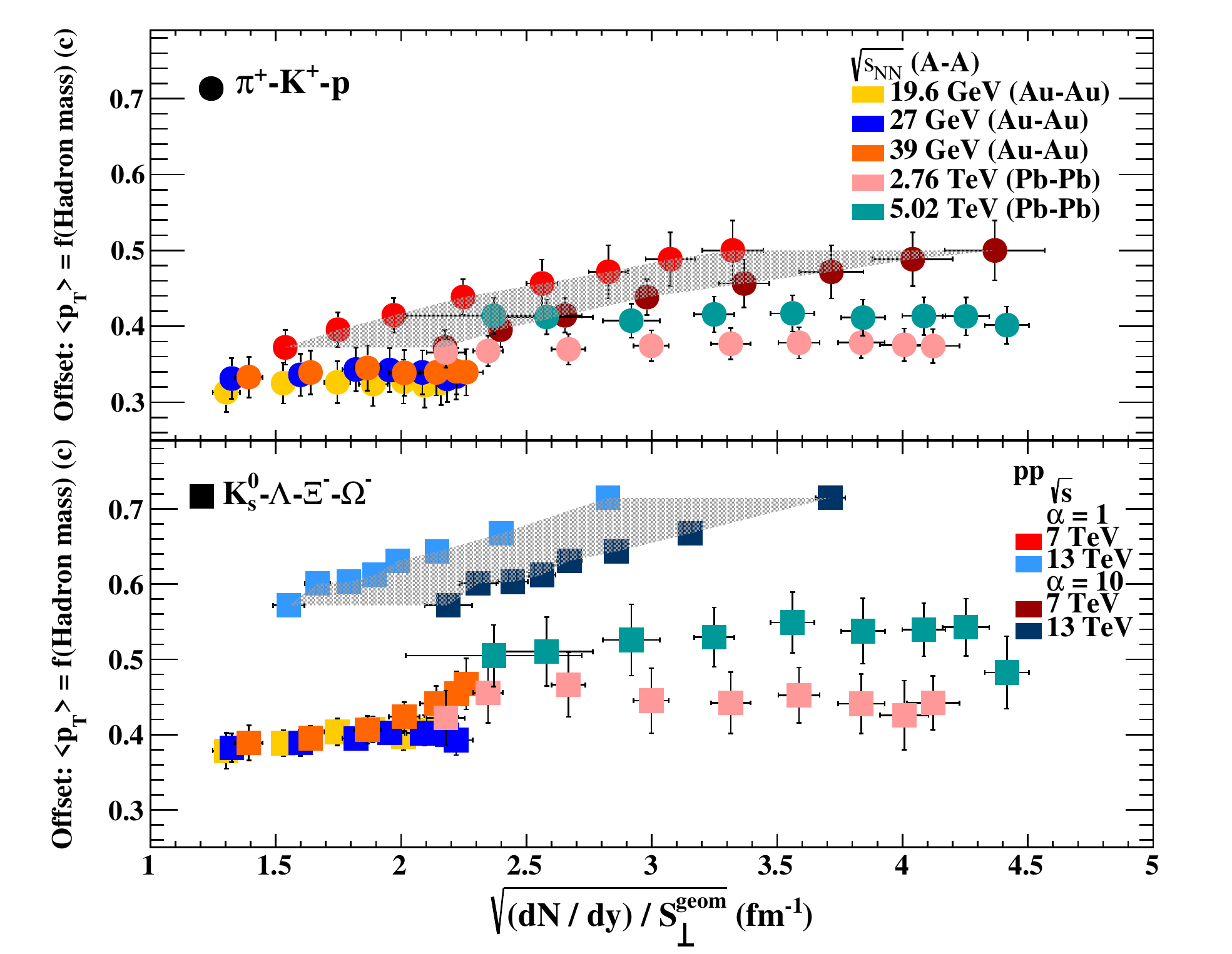}
\caption{$\sqrt{(dN/dy)/S_\perp}$ dependence of the offset obtained from the linear fit of the $\langle p_T \rangle$-hadron mass dependence for $\pi^+$,$K^+$and $p$(a) and $K^0_s$,$\Lambda$,$\Xi^-$and $\Omega^-$(b).}
\label{fig-6}
\end{figure}

Another study, less biased by any model assumptions, is done in terms of the dependence of  $\langle p_T \rangle$ on the  hadron mass based on measured experimental data. Linear fits were performed separately for $\pi^+$, $K^+$ and $p$ and $K^0_s$, $\Lambda$, $\Xi^-$ and $\Omega^-$, for different centralities (\textit{A-A}) or multiplicity classes (\textit{pp}). The parameters of the linear fits are represented in Fig.\ref{fig-5} and Fig.\ref{fig-6}, respectively. The slope parameter gives an insight on the dynamics of the collision. Figure \ref{fig-5} shows that at LHC energies the slopes of the linear $\langle p_T \rangle$-hadron mass dependence for \textit{pp} collisions in the case of $\alpha=1$ scale rather well with the values determined for \textit{Pb-Pb} collisions as a function of $\sqrt{(dN/dy)/S_\perp}$. An offset is evidenced between LHC and BES energies for \textit{Pb-Pb} and \textit{Au-Au}, respectively. Systematically higher values are evidenced in the case of $\pi^+$, $K^+$ and $p$ (Fig.\ref{fig-5}a), than in the case of $K^0_s$, $\Lambda$, $\Xi^-$ and $\Omega^-$ (Fig.\ref{fig-5}b). The offset of the linear $\langle p_T \rangle$-hadron mass dependence (Fig.\ref{fig-6}), is independent on the geometrical variable for \textit{A-A} collisions. The offsets increase towards higher values of $\sqrt{(dN/dy)/S_\perp}$ for \textit{pp} collisions. As in the case of $T_{kin}$, systematically higher values are observed for strange and multi-strange hadrons in \textit{A-A} collisions which indicates an earlier hadronization of these species, corresponding to a higher temperature of the deconfined fireball. 

In conclusion, a comprehensive study in terms of geometrical scaling of different observables, strongly related to the collision dynamics, is presented. A rather good scaling is evidenced for all RHIC energies, and separately for LHC energies, for strange and multi-strange hadrons ($K^0_s$, $\Lambda$, $\Xi^-$, $\Omega^-$). The dependence of the $\langle p_T \rangle$ on the geometrical variable shows an excellent scaling between \textit{Pb-Pb} and \textit{pp} collisions (for $\alpha$=10) at LHC energies. Observables that describe the dynamics of the collision, $\langle \beta_T \rangle$ and the slope of the $\langle p_T \rangle$-hadron mass dependence show a very good scaling between \textit{A-A} and \textit{pp} collisions, for $\alpha=1$. The $T_{kin}$ parameter from BGBW fits of the $p_T$ spectra and the offset of the $\langle p_T \rangle$-hadron mass dependence have higher values for strange and multi-strange hadrons than for $\pi^+$, $K^+$ and $p$. This suggests an earlier freeze-out for hadrons that have strange quarks in composition, therefore a higher initial temperature of the fireball. 

\acknowledgments
This work was carried out under contracts sponsored by the Ministry of Research, Innovation and Digitization: RONIPALICE- 03/16.03.2020 (via Institute of Atomic Physics coordinating agency) and PN-19 06 01 03/2019.
%\vspace{-0.5cm}
%%%%%%%%%%%%%%%%%%%%%%

\end{document}